\documentclass[twoside,fleqn]{article}

\usepackage{epsf}
\usepackage{espcrc2}
\usepackage{amsmath}
\usepackage{amssymb}

\title{
Weak matrix elements for CP violation
	\thanks{Poster presented by W.~Lee${}^{\rm a}$, This work has
	been supported by DOE.}
        }
\author{T.~Bhattacharya\address{MS--B285, Group T-8, 
		Los Alamos National Lab, Los Alamos, New Mexico 87545, USA}, 
	G.T.~Fleming\address{Department of Physics,
		Ohio State University, Columbus, OH 43210, USA}, 
	G.~Kilcup${}^{\rm b}$,
	R.~Gupta${}^{\rm a}$, 
	W.~Lee${}^{\rm a}$ 
	and 
	S.~Sharpe\address{Department of Physics,
		University of Washington, Seattle, WA 98195, USA} 
       }
       
\begin{document}
\begin{abstract}
We present preliminary results of matrix elements of four fermion
operators relevant to the determination of $\epsilon$ and
$\epsilon'/\epsilon$ using staggered fermions.
\end{abstract}
\maketitle

\section{INTRODUCTION} \label{sec:intr}

%
%
%
%
%
%
%
%
%
%
%

%
%
To calculate the matrix elements relevant to CP violation in Kaon
decays it is important to use a lattice formulation which preserves
(some) chiral symmetry. In the case of $B_K$, $B_7^{(3/2)}$, and
$B_8^{(3/2)}$, the absence of chiral symmetry leads to mixing with
wrong chirality operators, which in turn leads to large discretization
errors. The problem is far more severe for $B_6^{(1/2)}$ due to mixing
with lower dimension operators. Two lattice formulations that respect
at least part of the continuum chiral transformations and hold promise
for these calculations are domain wall/overlap fermions and staggered
fermions. Each has its advantages and disadvantages. Renormalization
of operators in the domain wall/overlap formulation is small enough
that 1-loop calculations may be adequate, but the numerical
simulations are $\sim 100$ times more costly. Staggered simulations
are very efficient, but the 1-loop renormalization constants for the
simplest lattice transcription of operators are very large. The goal
of this project is to find an improved staggered formulation for which
perturbation theory is well-behaved.

Here we present preliminary estimates of $B_K$, $B_7^{(3/2)}$,
$B_8^{(3/2)}$, and $B_6^{(1/2)}$ using 140 quenched lattices ($16^3
\times 64$) at $\beta = 6.0$.  This numerical simulation is being done
on the QCDSP supercomputer at Columbia University. To facilitate
chiral extrapolations, we have used four values of quark mass: $a m_q
= 0.01,$ $0.02,$ $0.03,$ $0.04$. The results are for gauge invariant
staggered operators that lie in a $2^4$ hypercube for which the 1-loop
renormalization constants are now known~\cite{wlee:0}.
%


\section{$B_K$} 
\label{sec:BK}
\begin{figure}[!t]
\epsfxsize=\hsize\epsfbox{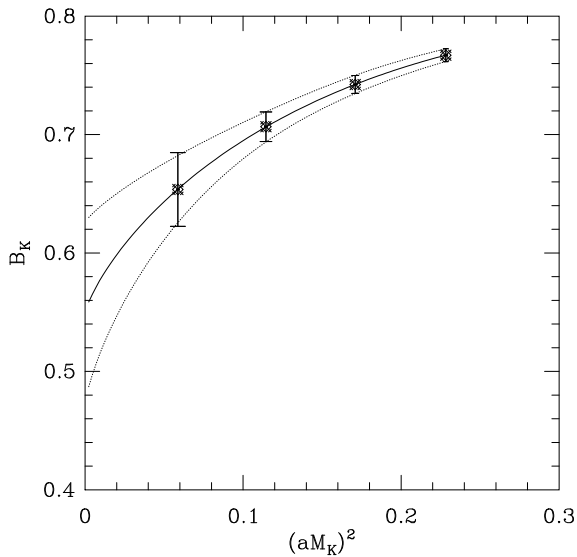}
\vskip -5mm
\caption{$B_K(\mu=\pi/a, NDR)$.}
\label{fig:b_k}
\vskip -4mm
\end{figure}

We have used the calculation of $B_K$ as a test of our programs. The
estimates shown in Figure~\ref{fig:b_k} agree with previous
calculations~\cite{kilcup:1,jlqcd:2,kilcup:0} and it is worth
mentioning that the 1-loop calculations done independently in
Ref.~\cite{wlee:0} reproduce the results given in
Ref.~\cite{jlqcd:1}. The figure shows a fit using the form suggested
by chiral perturbation theory, $c_0 + c_1 (a M_K)^2 + c_2 (a M_K)^2
\ln (a M_K)^2$, with $c_0 = 0.55(8)$, $c_1=0.06(25)$, and
$c_2=-0.60(38)$. Physical kaons correspond to $(a M_K)^2 \approx
0.06$.

%
%

\section{$B_7^{(3/2)}$ and $B_8^{(3/2)}$}
\label{sec:B7-B8}
\begin{figure}[!t]
\epsfxsize=\hsize\epsfbox{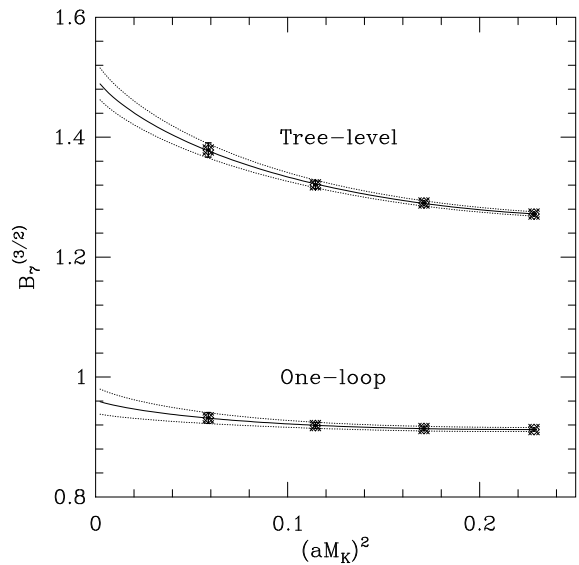}
\vskip -5mm
\caption{$B_7^{(3/2)}(\mu=\pi/a, NDR)$.}
\label{fig:b_7}
\vskip -4mm
\end{figure}

\begin{figure}[t]
\epsfxsize=\hsize\epsfbox{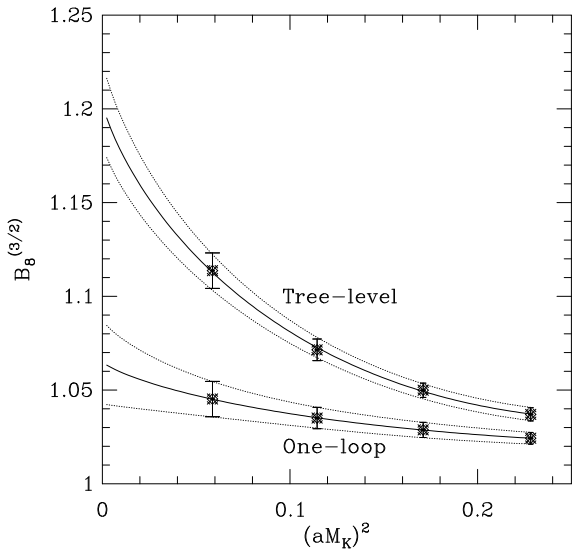}
\vskip -5mm
\caption{$B_8^{(3/2)}(\mu=\pi/a, NDR)$.}
\label{fig:b_8}
\vskip -4mm
\end{figure}

In Figures \ref{fig:b_7} and \ref{fig:b_8}, we compare the tree level
and one-loop results for $B_7^{(3/2)}$ and $B_8^{(3/2)}$ (defined
in~\cite{kilcup:1}). The 1-loop tadpole improved renormalization
constants for the gauge invariant operators were recently calculated 
in~\cite{wlee:0}. We match to the continuum NDR scheme at $q^* = \mu = \pi/a$.
The dominant contribution to $\langle O_8^{(3/2)} \rangle$ comes from
2-color trace staggered operator $[P \times P][P \times P]_{II}$,
which also dominates the vacuum saturation contribution. Thus, even
though the renormalization constants are large, there is a close
cancellation and the $B$ parameter receives a $\lesssim 10\%$ 1-loop
correction.  On the other hand $\langle O_7^{(3/2)} \rangle$ is
dominated by the 1-color contraction, and the renormalization
constants do not cancel. Consequently, even assuming $q^* = \pi/a$, 1-loop 
perturbation theory is unreliable.

These results can be compared against previous calculations done using
gauge non-invariant Landau gauge operators in
Ref.~\cite{kilcup:1}. The authors of Ref.~\cite{kilcup:1} found
significant systematic differences between results obtained using
smeared and unsmeared operators.  Choosing the same value of $\beta$
and $am_q$, we find that our results lie in between. 
%

%
%
%
%
%



%
%

\section{$B_6^{(1/2)}$} 
\label{sec:DeltaI=1/2}

Accurate estimates of $\epsilon'/\epsilon$ require measurements of
matrix elements of QCD and electromagnetic penguin operators, $\langle
O_6^{(1/2)} \rangle$ and $\langle O_8^{(3/2)} \rangle$ respectively,
in the $K \to \pi \pi$ transition.  Since these two contribute with
opposite sign, leading to a significant cancellation, both need to be
measured precisely in order to test whether the Standard Model
explains the observed size of ${\rm Re}(\epsilon'/\epsilon)$.

\begin{figure}[t]
\epsfxsize=\hsize\epsfbox{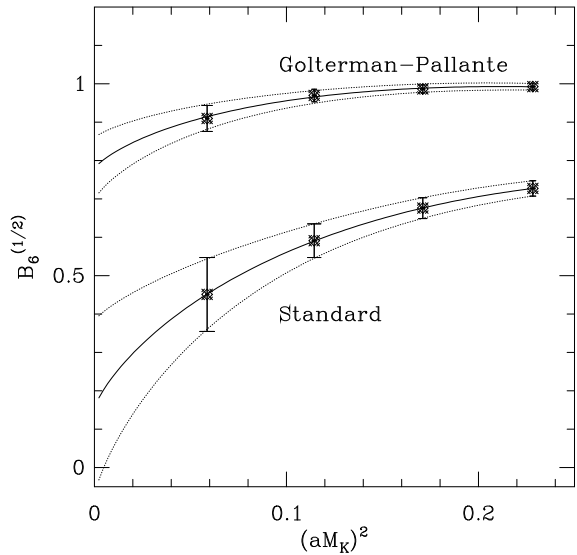}
\vskip -5mm
\caption{$B_6^{(1/2)}(\mu=\pi/a, NDR)$ 
	using the standard operator and 
	Golterman--Pallante operator.}
\label{fig:b_6}
\vskip -4mm
\end{figure}

Direct calculations of the $K \rightarrow \pi\pi$ amplitudes on the
lattice are difficult~\cite{testa:0}. The simplest
approach has been to assume chiral perturbation theory provides
accurate relations between $K \rightarrow \pi\pi$, $K\rightarrow \pi$
and $K \rightarrow 0$ amplitudes~\cite{bernard:0}. In this method,
which we use, the operator $O_6^{(1/2)}$ has three types of
contractions: eight, eye, and subtraction (to remove mixing with lower
dimension operators) \cite{kilcup:0,bernard:0}. Schematically,
\begin{eqnarray*}
B_6^{(1/2)} & = & 
	\frac{ \langle K^+ \mid O_6^{1/2} \mid \pi^+ \rangle }
	{ \mbox{Vac. Sat.} }
\\
O_6^{1/2} &=& O_6^{1/2} \mbox{\footnotesize(Eight)} 
	+ O_6^{1/2} \mbox{\footnotesize(Eye)} 
	+ O_6^{1/2} \mbox{\footnotesize(Sub)}
\end{eqnarray*}
In order to restrict the subtraction term to the single dimension four
operator $O_{\rm sub} \equiv (m_d-m_s){\bar s} \gamma_5 d + (m_d+m_s){\bar
s} d $, we need to work with degenerate $s$ and $d$ quarks. For $m_s =
m_d$ we can determine the coefficient $\alpha$ of $O_{sub}$ by calculating the
derivative of the $K \rightarrow 0$ amplitude with respect to the
strange quark mass.
This derivative introduces two types of diagrams: (i) each strange
quark propagator is replaced by its derivative, (ii) a disconnected
diagram that arises from the differentiation of the fermion
determinant. 
In principle, both terms can be estimated in the quenched theory.
However, the disconnected contribution vanishes in the limit of $m_s =
m_d$.
Hence, our calculation of $\alpha$ does not include the disconnected
contributions and is thus directly comparable to those of
Refs.~\cite{kilcup:0,norman:0,jlqcd:0}, which do not discuss such
contributions.
The results are shown in Figure \ref{fig:b_6}, labeled as ``standard''.
%
%
%
The details will be presented in Ref.~\cite{wlee:1}.

Recently, Golterman and Pallante have pointed out a subtlety
associated with the quenched approximation using (partially) quenched
chiral perturbation theory \cite{golterman:0}. They show that the
usual quenched operators lead to matrix elements which have a
different chiral expansion from that in the continuum.  They propose
an alternative quenched operator which does not have these
problems. In practice this amounts to dropping certain contractions in
the ``eye'' and ``subtraction'' diagrams.


We demonstrate in Figure \ref{fig:b_6} the difference in $B_6$ using
the standard operator and that suggested by Golterman and
Pallante. The 1-loop renormalization constants in the NDR scheme are
from Ref.~\cite{wlee:0} and \cite{sharpe:0}. Our calculations show
that the Golterman-Pallante operator enhances $B_6$ by almost a factor
of two, which in turn would significantly increase the value of
$\epsilon'/\epsilon$ compared to results given in
~\cite{kilcup:0,jlqcd:0}. Thus the quenched approximation remains the
most significant drawback of such calculations, which we hope to
address in the future.
%


%
%

We thank N.~Christ, G.~Liu, R.~Mawhinney and L.~Wu for their support
of this project and assistance with simulations on the Columbia
supercompter, QCDSP, on which this calculation was performed.

%
%
%

%
%
%
%
%

%
%
%
%
%

%
%
%
%

%
%
%
%

\begin{thebibliography}{}
%
\bibitem{wlee:0} W.~Lee, 
	Phys. Rev. {\bf D64} (2001) 054505.
%
\bibitem{kilcup:1} G.~Kilcup, R.~Gupta, and S.~Sharpe,
	Phys. Rev. {\bf D57} (1998) 1654.
%
\bibitem{jlqcd:2} S. Aoki, {\it et. al.}, 
	Phys.~Rev.~Lett. {\bf 80} (1998) 5271. 
%
\bibitem{kilcup:0} D.~Pekurovsky and G.~Kilcup, 
	Phys. Rev. {\bf D64} (2001) 074502.
%
\bibitem{jlqcd:1} N.~Ishizuka and Y.~Shizawa,
	Phys. Rev. {\bf D49} (1994) 3918.
%
\bibitem{testa:0} C.~Lin, {\it et.~al.}, hep-lat/0104006.
%
\bibitem{bernard:0} C.~Bernard, {\it et.~al.}, 
	Phys. Rev. {\bf D32} (1985) 2343.
%
\bibitem{norman:0} T.~Blum, {\it et.~al.},
	hep-lat/0110075.
%
\bibitem{jlqcd:0} J.~Noaki, {\it et.~al.},
	hep-lat/0108013.
%
\bibitem{wlee:1} W.~Lee, {\it et.~al.}, in preparation.
%
\bibitem{golterman:0} M.~Golterman, {\it et.~al.}, hep-lat/0108010.
%
\bibitem{sharpe:0} S.~Sharpe and A.~Patel,
	Nucl. Phys. {\bf B417} (1994) 307.
%
%
%
%
%
\end{thebibliography}
\end{document}